\documentclass[twocolumn]{aastex631}
\usepackage{amsmath,amssymb,graphicx,booktabs}
\hypersetup{hypertexnames=false}

\newcommand{\dm}{\delta_{\rm m}}

\shorttitle{Intermediate angle alignment}
\shortauthors{Lazarian}
\begin{document}
\title{Grain Alignment at Intermediate Angles: Frequency-Dependent Magnetic Relaxation}

\author[0000-0002-7336-6674]{A. Lazarian}
\affiliation{Department of Astronomy, University of Wisconsin--Madison, Madison, WI 53706, USA}

\begin{abstract}
Frequency-dependent magnetic dissipation can induce an incomplete alignment regime at intermediate-$J$ states in radiative torque (RAT) alignment, with similar effects expected for mechanical torque (MET) alignment. We demonstrate that this regime can be stationary or exhibit sustained oscillations of the angular momentum and its inclination to the magnetic field. Both types of attractors can prevent grains from reaching high-$J$ rotation, thereby reducing grain alignment for strong radiation fields. Trapping at stationary intermediate-$J$ attractors can delay the grain disruption for thousands of damping times, exceeding the lifetime of dust in many astrophysical settings. Using published Astrodust extinction efficiencies, we show that changes in alignment populations can produce the decrease of polarization with the increase of radiation intensity without disrupting grains. Apart from magnetic studies, the new state of grain rotation should be accounted for understanding grain physics, including grain disruption, coagulation, evolution of grain mantles and grain chemistry.
\end{abstract}

\section{Introduction}\label{sec:intro}
Polarization by aligned dust grains is an indispensable tracer of magnetic fields in the interstellar medium (ISM) and circumstellar environments \citep{Hildebrand00}. It acts as a foreground for studies of the cosmic microwave background \citep{2020A&A...641A...4P}. Interpreting it requires understanding how irregular grains acquire preferred alignment \citep{Andersson2015}. This enduring problem has drawn contributions from many of the founders of modern astrophysics. \citet{DavisGreenstein1951} identified paramagnetic dissipation as an important alignment mechanism, \citet{Gold1952} considered stochastic gaseous bombardment for mechanical alignment, while \citet{Purcell1979} and \citet{SpitzerMcGlynn1979} introduced the physics of suprathermal rotation and crossovers as the ways of enhancing paramagnetic alignment. Uncompensated spin-up torques \citep{Purcell1979} and enhanced dissipation from magnetic inclusions \citep{JonesSpitzer1967} were also recognized as important components of grain alignment. These processes established the intimate connection between grain magnetism, rotation, and polarization.

\citet{DolginovMytrophanov1976} recognized that anisotropic radiation can exert systematic torques on irregular grains, but quantitative torque calculations were not yet available at that time. The numerical calculations of \citet{DW96,DW97}, including the interstellar radiation spectrum and describing grain phase trajectories, changed that situation, introducing the mechanism to the community. The analytical model of RATs introduced by \citet{LH07} explained the origin of low- and high-angular-momentum attractors. \citet{HL08} clarified the effects of thermal wobbling, grain flips, and the role of gas bombardment. The RAT predictions have gained substantial observational support \citep[reviewed by][]{Andersson2015}, making the RAT theory the founding block of the grain alignment paradigm. 

\citet{LH08mag} proposed that magnetic inclusions could greatly enhance RAT alignment. The quantitative magnetic RAT (MRAT) framework was elaborated by \citet{HL16} forming the basis for modern understanding of grain alignment. Stronger magnetic dissipation enlarges the parameter space for perfect alignment at high-$J$ attractors. RATs can also spin grains to high angular velocities that can disrupt grains via centrifugal stress \citet{Hoang2019,Hoangetal2019}. This makes RATs important for determining the size distribution of dust grains in various astrophysical environments. 

\citet{LH2021} related alignment and disruption and showed that increasing the radiative torque can delay \emph{slow} disruption by retaining grains near low-$J$ attractors. Exploring the time response of alignment demonstrated the prospects for time-domain polarization studies. Combined with \citet{LH2019} analysis of magnetic precession and frequency-dependent grain magnetism, it opened a way to use observations for exploring grain magnetic properties.  

The present study asks whether magnetic relaxation can weaken during spin-up \emph{before} grains reach their disruption threshold. We investigate how the competition between RATs and magnetic torques related to magnetic dissipation changes the grain dynamics. We show that, in addition to the traditional low- and high-$J$ attractors, grains can possess intermediate-$J$ attractors associated with stationary or oscillatory rotation. We explore the consequences of this new regime for grain alignment, rotational disruption, and polarization in illuminated clouds. 

This study is important both for the alignment and disruption by RATs and Mechanical Torques (METs). The latter mechanism was introduced in \cite{LH2007met} where it was shown that the drift of irregular dust grains in respect to gas induces regular torques (see numerical studies in \cite{HoangChoLazarian2018, DasWeingartner2016}). Those arise from the intrinsic helicity of individual irregular grains, i.e., they are similar to RATs. A companion paper on METs (Lazarian 2026) establishes that METs for a statistical ensemble of grains are similar to the Analytical Model (AMO) successfully introduced for RATs in \cite{LH07}. Thus, most of the results we obtained in this paper for RATs can be applied to METs.

We present the physics of RAT alignment in the presence of magnetic relaxation in Section~\ref{sec:physics}. The physics of partial grain alignment at reduced rotational rates is presented in Section~\ref{sec:AMO}. Section \ref{sec:disr} is devoted to modification of grain disruption, and the effect of decrease of polarization with the increase of radiation intensity is demonstrated in   Section~\ref{sec:pol}.  Section~\ref{sec:disc} provides a discussion of results. In particular, big picture of our advance is presented in \S \ref{sec:big_picture}. Our summary is provided in Section \ref{sec:summary}. 

\section{Basics of Grain Dynamics}\label{sec:physics}
\subsection{RATs: alignment, disruption and low-$J$ survival}
\begin{figure}[t]
\centering
\includegraphics[width=\columnwidth]{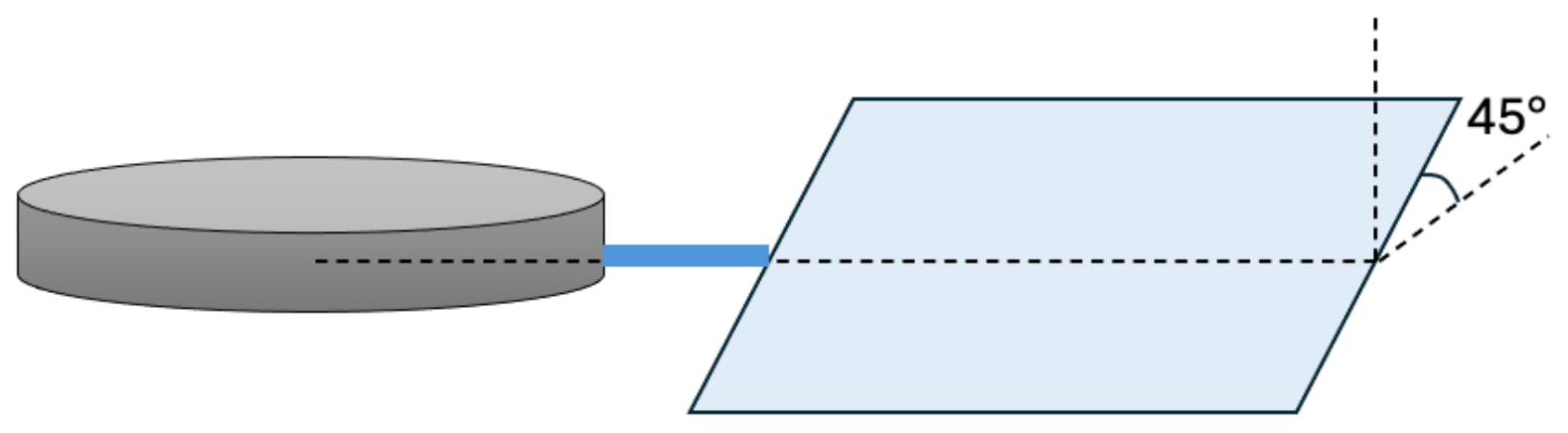}
\caption{Analytical model of radiative torques (AMO; \citealt{LH07}). A reflecting mirror attached to an oblate grain introduces helicity. The schematic illustrates a mirror inclination of $45^\circ$, while the reference numerical model below uses $55^\circ$.}
\label{fig:amo}
\end{figure}

Let $\boldsymbol J$ denote grain angular momentum, $J=|\boldsymbol J|$, and $\boldsymbol B$ the magnetic field. For a grain of effective radius $a$, the spectrum-averaged RAT is obtained by integrating the signed torque-efficiency vector over the anisotropic illuminating radiation:
\begin{equation}
 \boldsymbol\Gamma_{\rm RAT}=\pi a^2\int d\lambda\,\gamma_\lambda u_\lambda
 \frac{\lambda}{2\pi}\boldsymbol Q_\Gamma(a,\lambda).
 \label{eq:rat-spectrum}
\end{equation}
Here $\lambda$ is wavelength, $u_\lambda$ is radiation energy density per unit wavelength, $\gamma_\lambda$ is the anisotropy fraction, and $\boldsymbol Q_\Gamma$ is the dimensionless torque efficiency. Irregular grains can possess helicity. The Analytical Model (AMO) \citep{LH07} describes the RATs using a helical toy model with reflecting mirror attached to a grain body. This model reproduces important features of RATs of irregular grains \cite{LH07, Herranen2019}. Figure~\ref{fig:amo} illustrates the AMO geometry.

Let $\hat{\boldsymbol J}=\boldsymbol J/J$ and let $\xi$ be the angle between $\boldsymbol J$ and $\boldsymbol B$, $\psi$ the angle between the radiation direction $\hat{\boldsymbol k}$ and $\boldsymbol B$, and $\phi$ the Larmor precession phase. We write $\boldsymbol\Gamma_{\rm RAT}=M\boldsymbol Q$, where $M$ is a dimensional torque normalization proportional to the radiation intensity and $\boldsymbol Q$ is the corresponding dimensionless torque vector. The spin-up and aligning components are defined by
\begin{equation}
 H=\boldsymbol Q\cdot\hat{\boldsymbol J},\qquad
 F=\boldsymbol Q\cdot\hat{\boldsymbol e}_{\xi},\qquad
 \hat{\boldsymbol e}_{\xi}=\partial\hat{\boldsymbol J}/\partial\xi.
\end{equation}
Positive $H$ increases $J$, whereas positive $F$ increases $\xi$. The remaining component, along the azimuthal direction about $\boldsymbol B$, contributes to precession. We use angle brackets on $F$ and $H$ to denote averaging over $\phi$ at fixed $\xi$ and $\psi$. 

Both RATs and the magnetic field induce grain precession. When RAT-induced precession dominates, the preferred alignment axis can be $\hat{\boldsymbol k}$. The RAT torque geometry then corresponds to zero angle between the radiation and precession axes, but the magnetic relaxation terms must be reconsidered. Here we assume magnetic-field-dominated precession.  

The commonly used analytical approximation to the LH07 AMO torque functions for perfect internal alignment, i.e. with $\hat{\boldsymbol J}$ along the axis of grain maximal moment of inertia, is
\begin{align}
 Q_{e1}(\Theta)&=Q_{e1}^{\max}(5\cos^2\Theta-2)/3,\\
 Q_{e2}(\Theta)&=Q_{e2}^{\max}\sin 2\Theta,
\end{align}
where $\cos\Theta=\cos\xi\cos\psi+\sin\xi\sin\psi\cos\phi$. The scattering-frame axes are $\hat{\boldsymbol e}_1=\hat{\boldsymbol k}$, $\hat{\boldsymbol e}_2=(\hat{\boldsymbol J}-\cos\Theta\,\hat{\boldsymbol k})/\sin\Theta$, and $\hat{\boldsymbol e}_3=\hat{\boldsymbol e}_1\times\hat{\boldsymbol e}_2$. Here $Q_{ei}=\boldsymbol Q\cdot\hat{\boldsymbol e}_i$ and $Q_{ei}^{\max}$ denotes the maximum of the corresponding torque function. The ratio $q^{\max}=Q_{e1}^{\max}/Q_{e2}^{\max}$ characterizes the torque geometry. More generally, for AMO, the scattering-frame torque functions depend on the mirror inclination $\alpha$. In \cite{LH07} $\alpha=45^\circ$ was adopted, but to approximate the response of grains of different shapes and at different wavelengths $\alpha$ can be adjusted \citep{Herranen2019}.
In this paper, we adopt the reference mirror inclination $\alpha=55^\circ$ and radiation--field angle $\psi=30^\circ$.

RATs can spin grains up or down depending on orientation. The thermal angular-momentum scale for gas excitation is $J_{\rm th}=\sqrt{Ik_{\rm B}T_g}$, where $I$ is the maximum principal moment of inertia, $k_{\rm B}$ is Boltzmann's constant, and $T_g$ is the gas temperature. Internal thermal wobbling is governed instead by the grain temperature $T_d$ and the principal moments of inertia \citep{LazarianDraine1999Flips}. At angular momenta of order $J_d=(Ik_{\rm B}T_d)^{1/2}$ or below, thermal wobbling of the grain body relative to $\boldsymbol J$ reduces the effective spin-up and aligning RAT components upon averaging over the internal thermal distribution \citep{HL08}. As $J/J_d$ increases, internal alignment improves and the averaged torques approach their suprathermal forms. Thermal flips \citep{LazarianDraine1999Flips} reverse the orientation of the grain body relative to $\boldsymbol J$ and also modify the effective torques. A crossover does not necessarily produce sustained spin-up, and the averaged low-$J$ attractor is displaced from $J=0$ by fluctuations. After a flip, the thermally averaged RAT may continue to decelerate the grain on the opposite branch. Repeated low-$J$ returns therefore remain possible creating trapping of grains at low-$J$ attractors for many damping times while gaseous bombardment tries to get them out \citep{LH07, LH2021, Hoang2025General}. 

For efficient internal alignment the rotation frequency is $\omega=J/I$. The centrifugal grain disruption was introduced in \citep{Hoang2019,Hoangetal2019}. The condition for it to happen is
\begin{equation}
 \omega\gtrsim\omega_{\rm disr}\simeq\frac{2}{a}
 \left(\frac{S_{\max}}{\rho}\right)^{1/2},
 \label{eq:disr}
\end{equation}
with tensile strength $S_{\max}$ and grain mass density $\rho$. We define $J_{\rm disr}=I\omega_{\rm disr}$ for perfect internal alignment. A large nominal RAT spin-up rate does not itself imply disruption: the actual grain must enter a trajectory that reaches angular velocity given Eq.~(\ref{eq:disr}). Magnetic response stabilizes high-$J$ attractors \citep{LH08mag} and therefore affects both the probability of reaching high $J$ and the survival of grains en route.

When the stable high-$J$ attractor is missing, the low-$J$ population is weakly aligned because thermal wobbling and flipping weaken the connection between the grain axis and $\boldsymbol J$ \citep{HL08,LH2021}. Nevertheless, a trajectory released by gas impacts can pass near a high-$J$ repeller and reach $J_{\rm disr}$ even though no stable high-$J$ state exists. As demonstrated by \citet{LH2021}, stronger radiation can make these exceptional excursions less probable, since the systematic torques return perturbed grains more efficiently to low $J$.

\subsection{Enhanced magnetic response and frequency dependence of magnetic relaxation}

Superparamagnetic inclusions increase magnetic relaxation and can stabilize aligned high-$J$ attractors absent in an inclusion-free grain \citep{LH08mag,HL16}. Grains whose trajectories enter their domains then approach high $J$ directly, yielding strong alignment without a long low-$J$ residence. If the RAT spin exceeds $J_{\rm disr}$, this is also the fast-disruption channel of \citet{LH2021}. The same enhanced dissipation can make stochastic low-$J$ releases more likely to end at high $J$, further shortening disruption times. In this paper we discuss what happens if magnetic relaxation weakens as the grain rotational velocity increases. 

A spinning grain with a magnetic moment undergoes precession about the ambient magnetic field. For the magnetic moment generated through the Barnett effect $\mu_{\rm Bar}$  \citep{DolginovMytrophanov1976}, the approximate precession frequency is $\Omega_B=\mu_{\rm Bar}B/J$. In the low-frequency Barnett approximation $\mu_{\rm Bar}\propto \chi(0)\omega$, whereas the dissipation that tends to align $\boldsymbol J$ with $\boldsymbol B$ depends on the \emph{imaginary} susceptibility, $K(\omega)=\chi_2(\omega)/\omega$. Here $\chi(0)$ is the static susceptibility and $\chi_2$ is the imaginary part of the susceptibility. The characteristic magnetic relaxation time and its ratio relative to rotational damping 
are (\cite{JonesSpitzer1967}):
\begin{equation}
 \tau_{\rm mag}=\frac{I}{VB^2K(\omega)},\qquad
 \dm(\omega)=\frac{\tau_{\rm damp}}{\tau_{\rm mag}},
 \label{eq:mag}
\end{equation}
where $I$ is the moment of inertia about the maximum-inertia axis, $V$ is grain volume, and $B=|\boldsymbol B|$. The total rotational damping time $\tau_{\rm damp}$ includes gas damping ($\tau_{\rm gas}$) and infrared damping ($\tau_{\rm IR}$):
\begin{equation}
 \tau_{\rm damp}^{-1}=\tau_{\rm gas}^{-1}+\tau_{\rm IR}^{-1},\qquad F_{\rm IR}=\tau_{\rm gas}/\tau_{\rm IR}.
\label{eq:damping-definition}
\end{equation}
Thus $\tau_{\rm gas}=(1+F_{\rm IR})\tau_{\rm damp}$ and the infrared fraction of the total damping rate is $f_{\rm IR}=F_{\rm IR}/(1+F_{\rm IR})$. The rotational frequency is approximated by $\omega=J/I$ when internal alignment is efficient. 

Strong magnetic precession and strong magnetic \emph{dissipation} are distinct requirements. In particular, a decrease in $K(\omega)$ does not necessarily decrease precession about $\boldsymbol B$ \citep{LH2019}.

Superparamagnetic inclusions can enhance low-frequency $K$, but finite cluster and spin-relaxation times suppress dissipation at sufficiently fast rotation. Ferromagnetic and ferrimagnetic inclusions can also enable enhanced spin--lattice dissipation in the surrounding paramagnetic material. In the model adopted here, the ordinary paramagnetic contribution remains when the inclusion-induced enhancement declines. Figure~\ref{fig:mag} illustrates the frequency dependence of $K(\omega)$ using the model response laws specified in Appendix~\ref{app:maglaws}. The corresponding magnetic-relaxation parameter $\delta_m(\omega)$ follows from Equation~(\ref{eq:mag}). We require $\Omega_B$ to dominate radiative precession and secular alignment and assume efficient grain-body alignment with $\boldsymbol J$ outside low $J$. The effect of rotation on internal alignment is considered in Section~\ref{sec:sizedep}.

\begin{figure}[t]
\centering\includegraphics[width=\columnwidth]{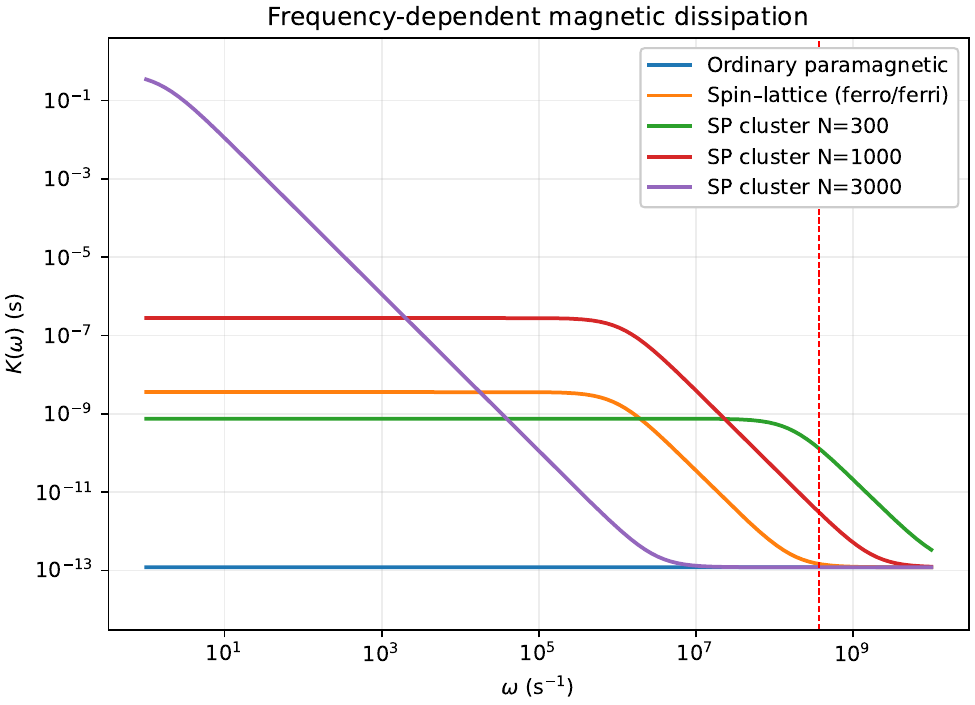}
\caption{Frequency dependence of the dissipative magnetic response $K(\omega)=\chi_2(\omega)/\omega$ for ordinary paramagnetic material, a paramagnetic grain with an additional inclusion-induced spin--lattice contribution, and illustrative superparamagnetic clusters with the inclusion sizes indicated in the legend. Response laws are specified in Appendix~\ref{app:maglaws}, following the magnetic-relaxation treatment in \citet{LH2019}. The vertical dashed line marks $\omega_{\rm disr}\simeq3.7\times10^8\,{\rm s}^{-1}$ for $a=0.1\,\mu{\rm m}$, $S_{\max}=10^7\,{\rm erg\,cm}^{-3}$, and $\rho=3\,{\rm g\,cm}^{-3}$ [Equation~(\ref{eq:disr})]. The enhanced responses approach the ordinary paramagnetic curve as the inclusion-related contributions decline.}\label{fig:mag}
\end{figure}

The magnetic response relevant to a particular grain is evaluated at its \emph{actual} rotation frequency. Low-frequency dissipation need not extend to high frequencies. If it falls before the grain reaches high-$J$ state, the grain finds itself in the situation of no high-$J$ option. Its trajectory should move back to low-$J$, but as the rotational frequency decreases, the magnetic torque favoring high-$J$ alignment recovers. The grain finds itself in a position of Buridan's ass settling at intermediate angle and intermediate $J$ at a new type of attractor that we quantitatively discuss further.

\section{Grain as a Buridan ass: From high-$J$ capture to intermediate-$J$ trapping}\label{sec:AMO}
\subsection{New attractors at intermediate $J$}
Let $U$ denote the radiation energy density in units of the standard interstellar radiation energy density $u_{\rm ISRF}$, with fixed spectral shape and anisotropy. For the reference torque geometry we define the nominal RAT spin by
\begin{equation}
 J_{\rm RAT}=M\langle H(0,\psi)\rangle\tau_{\rm damp},
 \label{eq:jrat_scale}
\end{equation}
assuming positive spin-up at $\xi=0$. This is the spin obtained by balancing RATs against rotational damping, regardless of angular stability. We choose $U_0$ such that $J_H\equiv J_{\rm RAT}(U_0)=50J_{\rm th}$ in the reference benchmark. With $u=U/U_0$, $M=uM_0$, where $M_0=M(U_0)$, and fixed damping and torque geometry, $J_{\rm RAT}=uJ_H$. Define $\omega_H=J_H/I$ and $j=J/J_H$, $j_{\rm th}=J_{\rm th}/J_H$, $j_{\rm disr}=J_{\rm disr}/J_H$, and $t'=t/\tau_{\rm damp}$. The normalized torque functions are $f=M_0\tau_{\rm damp}\langle F\rangle/J_H$ and $h=M_0\tau_{\rm damp}\langle H\rangle/J_H$. We write $\delta_m(j)=\delta_m(\omega=j\omega_H)$. For a magnetic response with turnover frequency $\omega_c$, we define $J_c=I\omega_c$ and $j_c=J_c/J_H=\omega_c/\omega_H$. Dots denote derivatives with respect to $t'$. The reduced equations are \citep{LazarianHoang2008,HL16}
\begin{align}
 \dot j &= u h(\xi)-j\left[1+\dm(j)\sin^2\xi\right],\label{eq:dj}\\
 \dot\xi &=\frac{u f(\xi)}{j}-\dm(j)\sin\xi\cos\xi.\label{eq:dxi}
\end{align}
In the reference torque geometry, grains without magnetic relaxation predominantly return to low $J$, whereas constant strong magnetic dissipation stabilizes aligned high-$J$ rotation. If $\delta_m(j)$ decreases during spin-up, the aligned high-$J$ equilibrium can become unstable, and the angular and spin balances may instead yield a stable off-axis intermediate-$J$ equilibrium ($\dot j=\dot\xi=0$) even when high-$J$ alignment is unstable. The equilibria of Equations~(\ref{eq:dj}) and (\ref{eq:dxi}), including attractors, repellers, and saddle points, will be referred to collectively as stationary points \citep{DW97}.

\begin{figure*}[t]
\centering\includegraphics[width=0.96\textwidth]{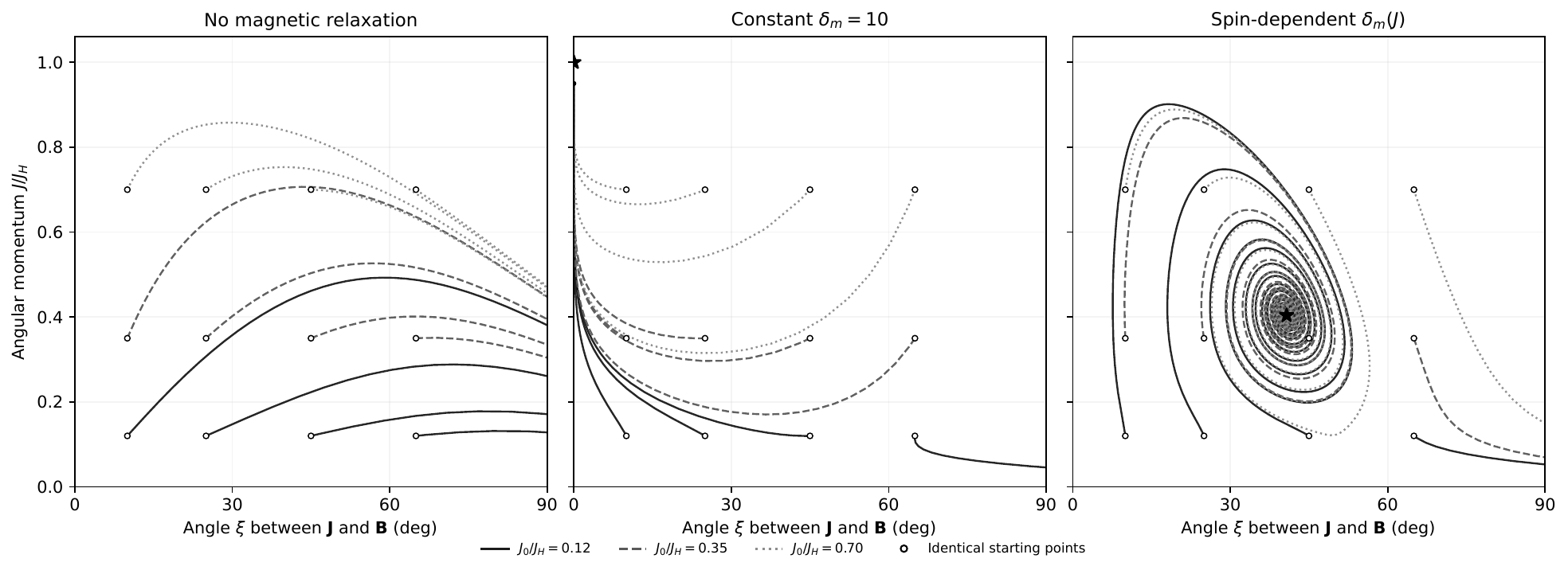}
\caption{Evolution of grain rotation and orientation for the same twelve initial states in three magnetic-response models: no magnetic relaxation (left), constant $\delta_m=10$ (middle), and $\delta_m(J)=10/[1+(J/0.2J_H)^2]$ (right). Open circles mark the common starting states. With no magnetic relaxation the grains return toward low $J$; strong constant relaxation leads most to high $J$; and frequency-dependent relaxation allows an intermediate-$J$ attractor (star).}\label{fig:trajectories}
\end{figure*}

\begin{figure}[t]
\centering
\includegraphics[width=\columnwidth]{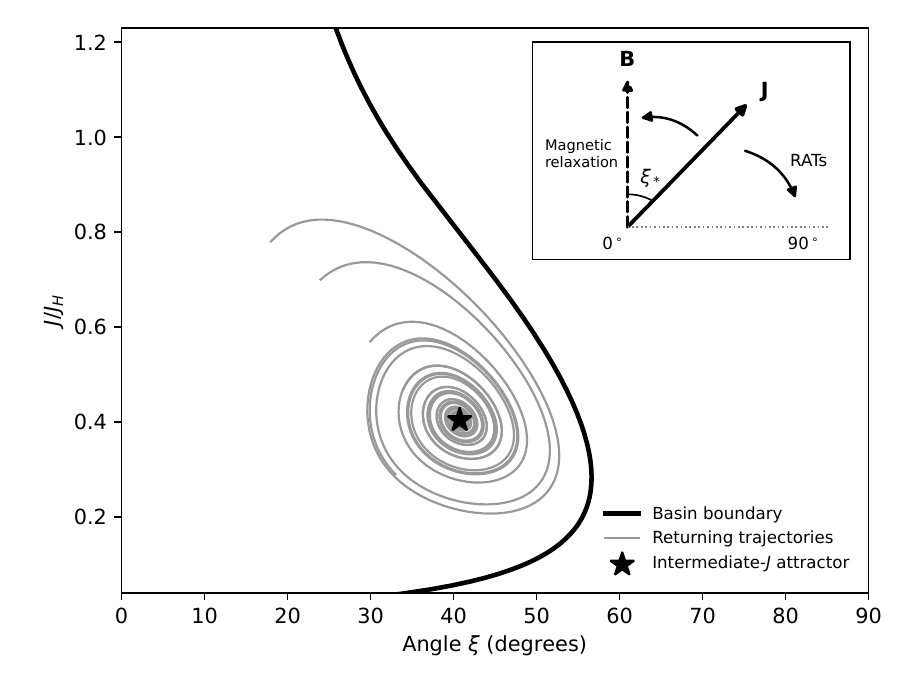}
\caption{Trajectories near the intermediate-$J$ attractor for the
reference frequency-dependent magnetic-response model. The basin of
attraction is the region in the $(J,\xi)$ plane containing initial
states from which grains evolve toward the attractor in the absence of
random torques. The black curve marks its boundary. The star marks the
attractor, and gray curves show trajectories approaching it. The inset
illustrates the competition between magnetic relaxation, which drives
$\boldsymbol{J}$ toward the magnetic-field direction, and RATs, which
drive it toward larger inclination. At the equilibrium inclination
$\xi_*$, these angular effects balance.}
\label{fig:ir_basin}
\end{figure}

The intermediate-$J$ attractor has a simple physical interpretation in terms of the competition between radiative and magnetic alignment. For the torque geometry considered, RATs drive grains toward larger inclinations and ultimately toward low-$J$ attractor. On the contrary, magnetic relaxation is associated with magnetic torque that drives {\bf J} to the magnetic-field direction corresponding to high-$J$ attractor, where RAT spin-up is more efficient. As the grain spins faster, the magnetic torque weakens, allowing RATs to increase the inclination and thereby reduce the spin-up torque. Conversely, a decrease in angular momentum/velocity enhances magnetic torque, favoring smaller inclinations and more efficient radiative spin-up. This feedback establishes an equilibrium at intermediate angular momentum and finite inclination, when the aligned high-$J$ stationary point is unstable due to fast rotation. Figure~\ref{fig:trajectories} compares identical initial conditions for weak, constant-strong, and frequency-dependent relaxation; only the last produces the illustrated intermediate attractor.\footnote{The intermediate-$J$ state is not limited to the chosen AMO mirror inclinations and radiation--field angles. For example, with the original LH07 mirror inclination of $45^\circ$, $\psi=30^\circ$, and $\delta_m(j)=10/[1+(j/0.2)^2]$, we obtain stable intermediate-$J$ attractors at $u=0.8$ and $0.9$. Their angular momenta are $j_*\simeq0.6$ and $0.5$, and their inclinations are approximately $26^\circ$ and $38^\circ$, respectively. The existence of these attractors is therefore not restricted to the adopted mirror inclination.} Figure~\ref{fig:ir_basin} shows the boundary separating trajectories that approach the intermediate-$J$ attractor from those outside its basin. 

\subsection{Stationary and oscillatory intermediate-$J$ alignment}\label{sec:oscillatory}
The feedback between radiative spin-up and frequency-dependent magnetic relaxation can also produce sustained oscillations of $J$ and $\xi$. As irradiation increases, the stationary intermediate-$J$ attractor can lose stability and be replaced by a stable periodic orbit. During the cycle, increasing $J$ weakens magnetic relaxation, allowing RATs to increase $\xi$ and reduce spin-up. Rotational damping then decreases $J$, restoring magnetic relaxation and more efficient radiative spin-up at smaller $\xi$. The oscillations arise from Equations~(\ref{eq:dj}) and (\ref{eq:dxi}) alone, without systematic body-fixed torques or random excitation.

\begin{figure}[t]
\centering\includegraphics[width=\columnwidth]{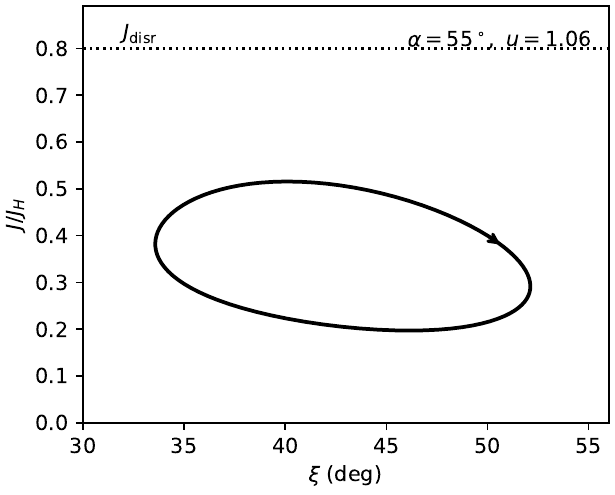}
\caption{Stable periodic intermediate-$J$ rotation for the reference AMO geometry ($\alpha=55^\circ$, $\psi=30^\circ$) at $u=1.06$, with $\delta_m(j)=10/[1+(j/0.2)^2]$ and no systematic body-fixed or stochastic torques. The arrow indicates the direction of evolution. The dotted horizontal line marks $J_{\rm disr}=0.8J_H$.}
\label{fig:oscillatory}
\end{figure}

Figure~\ref{fig:oscillatory} illustrates a periodic attractor for our reference AMO geometry ($\alpha=55^\circ$, $\psi=30^\circ$) with $\delta_m(j)=10/[1+(j/0.2)^2]$. The stationary point loses stability near $u=1.05$. At $u=1.06$, the grain oscillates between $0.2J_H$ and $0.52J_H$ with a period of about $4\tau_{\rm damp}$. Integration from different perturbations of the stationary point converges to the same periodic orbit. The time-averaged external alignment factor, $Q_J=\langle(3\cos^2\xi-1)/2\rangle_t$, is about 0.31. Thus grains captured by this attractor remain partially aligned and avoid disruption for $J_{\rm disr}=0.8J_H$. For oscillatory attractors, the maximum angular momentum reached during the cycle determines whether disruption occurs.

Systematic body-fixed torques, including Purcell torques \citep{Purcell1979} and infrared emission torques of irregular grains \citep{HoangLazarian2009Pinwheel}, can also shift the intermediate-$J$ stationary point and induce oscillatory rotation. For relatively cool grains rotating suprathermally, systematic infrared emission torques are generally weak. Neither type of systematic torque is required for the periodic attractor illustrated here.

\begin{figure}[t]
\centering\includegraphics[width=\columnwidth]
{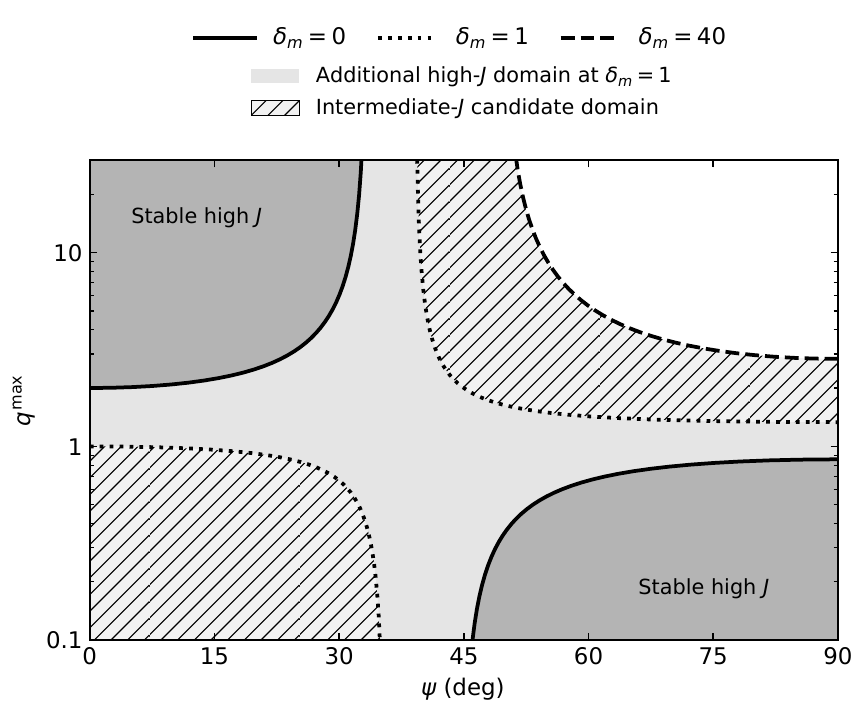}
\caption{Illustrative domain of intermediate-$J$ alignment in the analytical AMO, following the high-$J$ stability diagrams of \citet{LH07,HL16}. Solid, dotted, and dashed curves give the boundaries of high-$J$ stability for constant magnetic relaxation parameters $\delta_m=0$, 1, and 40, respectively, on the positive spin-up branch. Dark gray indicates high-$J$ stability without magnetic relaxation. Light gray indicates the additional region stabilized by $\delta_m=1$. The hatched region is the candidate domain of intermediate-$J$ alignment when the magnetic response decreases from 40 to 1 during spin-up. The reduced value is illustrative and depends on the grain torque geometry and rotation frequency.}
\label{fig:amo-regimes}
\end{figure}

The diagrams of \citet{LH07,HL16} characterize high-$J$ alignment in the $(q^{\max},\psi)$ plane. Figure~\ref{fig:amo-regimes} extends this approach to illustrate the domain of intermediate-$J$ alignment. We use the analytical AMO torque functions of Equations (61) and (62) of \citet{LH07} and the high-$J$ stability criterion of \citet{HL16}. On the positive spin-up branch, $\langle H(0,\psi)\rangle>0$, the critical magnetic relaxation parameter for high-$J$ stability is
\begin{equation}
 \delta_{m,{\rm crit}}(q^{\max},\psi)=
 \frac{\left.\partial\langle F(\xi,\psi)\rangle/\partial\xi\right|_{\xi=0}}
 {\langle H(0,\psi)\rangle}.
 \label{eq:amo_delta_crit}
\end{equation}
The aligned high-$J$ state is stable for $\delta_m>\delta_{m,{\rm crit}}$. As magnetic relaxation increases, the corresponding stable high-$J$ domain expands. For illustration, we take the low-frequency magnetic response $\delta_{m,0}=40$ and a reduced response $\delta_m=1$ at suprathermal rotation. The hatched region, $1<\delta_{m,{\rm crit}}<40$, identifies the domain where weakening magnetic relaxation during spin-up removes high-$J$ stability and gives rise to intermediate-$J$ alignment. The adopted reduced value $\delta_m=1$ is illustrative. The effective magnetic relaxation at intermediate rotation varies with $q^{\max}$ and $\psi$, which determine the balance between radiative and magnetic torques.

The shaded domain identifies where intermediate-$J$ alignment is possible, rather than distinguishing stationary and periodic attractors. To estimate their relative prevalence, we sampled this domain for $0.5\leq q^{\max}\leq4$ and $0^\circ\leq\psi\leq84^\circ$, using $\delta_m(J)=40/[1+(J/J_c)^2]$ with $J_c=0.15J_H$ and $J_H=J_{\rm RAT}$ for each geometry in this survey. Among the identified intermediate-$J$ states that remain below $J_{\rm disr}=0.8J_H$, approximately 80\% are stationary and 20\% are periodic. These proportions describe the sampled parameter space rather than grain occupation fractions. We find no coexistence of the two stable intermediate-$J$ states in the sampled models. Our goal here is to identify intermediate-$J$ alignment as a new regime of grain dynamics. For the remainder of this paper, we focus on stationary intermediate-$J$ attractors for the quantitative estimates of residence, disruption, and polarization. A detailed comparison of stationary and oscillatory intermediate-$J$ alignment will be presented elsewhere.

\subsection{Gas bombardment and slow transfer to high-$J$ states}\label{sec:stochastic}
Gas bombardment can release grains from low $J$, but repeated RAT-driven returns delay its escape from the low-$J$ attractor. This is the effect of trapping at low-$J$ attractors by RATs \citep{HL08,LH2021,Hoang2025General}. 

Our companion study of low-$J$ RAT and MET dynamics (Lazarian 2026) finds more frequent escape for irregular triaxial grains. For estimates in the regime where orientation-induced torque fluctuations facilitate escape, we estimate the mean waiting time $t_{\rm L}$ for an effective release into the resolved dynamics by an ad hoc prescription
\begin{equation}
 t_{\rm L}\sim 100\tau_{\rm gas}
 \left(\frac{J_{\rm RAT}}{50J_{\rm th}}\right)^{-1},
 \label{eq:low_wait_estimate}
\end{equation}
that roughly corresponds to the results of our low-$J$ dynamics investigation to be published elsewhere. We use this estimate for the illustrative disruption histories and do not infer a radiation dependence of the slow alignment time from it.

For the disruption histories in Figure~\ref{fig:ftimemodels}, we adopt an initial thermal population, with isotropic angular momentum directions and Maxwell-distributed magnitudes (each Cartesian component has dispersion $J_{\rm th}=0.02J_H$). Grains outside the low-$J$ core evolve immediately according to Equations~(\ref{eq:dj}) and (\ref{eq:dxi}), with stochastic torques. This allows the fast alignment channel of \citet{LH07} and the fast disruption channel of \citet{LH2021}. Grains initially inside, or subsequently returning to, the core ($j<0.025$) wait for an effective release. The waiting times are exponentially distributed with mean $100\tau_{\rm gas}$ at the adopted $u=1$. Each release starts at $j=0.06$ with an isotropic direction. Repeated releases and recapture give the slow alignment and disruption channels \citep{HL08,LH2021}. Frequency-dependent magnetic relaxation adds a third route, capture at intermediate $J$.

Infrared emission supplies half of the total damping in this comparison, so $\tau_{\rm damp}=0.5\tau_{\rm gas}$. We hold the nominal RAT spin and the dimensionless deterministic torque laws fixed at the reference normalization and reduce the stochastic diffusion coefficient to $0.505$ of its gas-only value [Equation~(\ref{eq:ir_diffusion_ratio})]. This isolates the effect of weaker stochastic excitation at fixed total damping. Time is expressed in units of $\tau_{\rm gas}$ in this figure.

The long delays arise because an individual release rarely delivers the grain directly to high $J$. If $p_H$ is the chance of successful transfer in one attempt and $t_{\rm cyc}$ is the mean duration of an attempt, including the low-$J$ wait and subsequent excursion, a useful estimate is
\begin{equation}
 \langle t_{\rm high}\rangle\sim \frac{t_{\rm cyc}}{p_H}.
 \label{eq:renewal_time}
\end{equation}
Thus, rare successful releases can delay alignment far beyond a damping time. Disruption may occur earlier if the grain crosses $J_{\rm disr}$.

\subsection{Escape from intermediate rotation and recurrent cycling}\label{sec:intermediate-escape}
Gas bombardment also controls the lifetime of intermediate-$J$ rotation. For the frequency-dependent response in Figure~\ref{fig:trajectories}, the attractor lies at $J/J_H\simeq0.4$ and $\xi\simeq41^\circ$. Most small perturbations decay as the grain returns to the attractor. Larger perturbations can carry the grain out of its basin of attraction and eventually back to low $J$.

The time to return to low rotation increases rapidly as the stochastic torques weaken. For instance, at $J_{\rm th}/J_H=0.02$, the median return time is about $50\tau_{\rm damp}$. Halving this ratio allows most grains to avoid returning to low $J$ throughout the calculation ($1000\tau_{\rm damp}$).

To quantify the onset of escape, we calculate the basin boundary in the $(J,\xi)$ plane (Figure~\ref{fig:ir_basin}). The mean time to first cross this boundary, rather than the time to reach low $J$, is given in Table~\ref{tab:ir_mfpt}.

Infrared emission can substantially increase residence times because it damps rotation while producing much smaller random angular-momentum increments than gas bombardment. To isolate this effect, we vary the fraction $f_{\rm IR}$ of rotational damping due to infrared emission at fixed total damping and deterministic torques. In our diffusion model, the ratio $\epsilon=0.01$ specifies that infrared emission contributes only $1\%$ as much angular-momentum diffusion per unit damping as gas collisions. Table~\ref{tab:ir_mfpt} shows that even a moderate infrared contribution substantially increases the residence time. When infrared emission accounts for three quarters of the damping, the enhancement reaches more than two orders of magnitude. The diffusion model and escape calculation are described in Appendix~\ref{app:ir-escape}.

\begin{table}[t]
\centering
\caption{Residence time inside the region of trajectories returning to the intermediate-$J$ attractor for
increasing infrared damping ($u=1$, $j_{\rm th}=0.02$,
$\epsilon=0.01$). Here $T$ is the mean first-exit time starting at the attractor and $T_0=T(f_{\rm IR}=0)$ is its value for gas damping alone.}
\label{tab:ir_mfpt}
\setlength{\tabcolsep}{7pt}
\begin{tabular}{ccc}
\hline\hline
$f_{\rm IR}$ & $T/\tau_{\rm damp}$ & $T/T_0$\\
\hline
0.00 & 39 & 1.0\\
0.25 & 78 & 2.0\\
0.50 & $\simeq 290$ & $\simeq 7$\\
0.75 & $\sim 10^4$ & $\sim 3\times10^2$\\
\hline
\end{tabular}
\end{table}

Periodic attractors can also decay through stochastic excitation. For three examples in the analytical AMO survey ($\delta_{m,0},q^{\max},\psi$ equal to $(40,1.3,10^\circ)$, $(40,1.5,32^\circ)$, and $(60,0.9,20^\circ)$), we start grains at random phases of the nondisruptive periodic orbit and use $f_{\rm IR}=0.5$ and $J_{\rm th}=0.02J_H$. The mean times to first leave the numerically estimated basin or reach the disruption threshold are about 14, 7, and $27\tau_{\rm damp}$, respectively. Stochastic excursions can therefore remove grains from periodic states on relatively short timescales. In these three examples most trajectories eventually cross the disruption threshold rather than return directly to the thermal core. Escape, return to low $J$, and disruption are distinct outcomes. The estimates are illustrative and do not imply that every periodic attractor is short-lived.

\subsection{Increasing irradiation can reduce grain alignment}
We describe grain alignment through the Rayleigh reduction factor
\begin{equation}
 R=\left\langle G(\cos\theta)G(\cos\xi)\right\rangle,
 \qquad G(x)=\frac{3x^2-1}{2},
 \label{eq:rayleigh-factor}
\end{equation}
where $\theta$ is the angle between the grain's maximum-inertia axis and $\boldsymbol J$, and the average is over the surviving grain population. For perfect internal alignment, $R=Q_J=\langle G(\cos\xi)\rangle$.

For a population of surviving grains, let $f_{\rm L}$, $f_{\rm S}$, $f_{\rm O}$, and $f_{\rm H}$ denote the fractions in low-$J$, stationary intermediate-$J$, oscillatory intermediate-$J$, and high-$J$ rotation. Their sum is unity. Taking the net alignment of low-$J$ grains to be negligible ($R_{\rm L}=0$), we have
\begin{equation}
 R=f_{\rm S}R_{\rm S}+f_{\rm O}R_{\rm O}+f_{\rm H}R_{\rm H}.
 \label{eq:alignment-regimes}
\end{equation}
Here $R_k$ denotes the mean alignment factor of grains in regime $k$. For oscillatory rotation the external alignment factor is averaged over the cycle, including its correlation with internal alignment when thermal wobbling matters. The observable alignment also depends on the fraction of grains surviving disruption and their optical cross sections.

To illustrate the effect of recurrent low-$J$ trapping, we follow the reference stationary and periodic models at $u=1$ and 1.06, respectively, for $500\tau_{\rm gas}$, using the thermal initial distribution, infrared damping, and release model of Section~\ref{sec:stochastic}, with mean release interval $100\tau_{\rm gas}/u$. We include repeated returns to low $J$, recapture, and disruption at $0.8J_H$. Taking $R=0$ at $j<0.06$ and perfect internal alignment at larger $j$, the survivor-weighted population alignment averaged over the final $100\tau_{\rm gas}$ is $R_{\rm pop}^{(\rm S)}\simeq0.06$ for the stationary model and $R_{\rm pop}^{(\rm O)}\simeq0.001$ for the periodic model (1200 histories per model). Thus appreciable alignment on a deterministic periodic orbit need not imply appreciable population alignment. For illustration, an equally weighted population with 80\% stationary and 20\% weakly aligned periodic realizations would reduce the stationary-only alignment estimate by about 20\%. This is modest compared with the decrease associated with the loss of high-$J$ alignment. We therefore retain the stationary population model for the illustrative polarization calculation.

Increasing irradiation can destabilize magnetically supported high-$J$ attractors [Equation~(\ref{eq:app_highj_instability})], redistributing grains toward low-$J$ and intermediate-$J$ states. The resulting decrease in alignment can precede rotational disruption.

A further reduction arises from changes in the intermediate-$J$ population itself. For the frequency-dependent reference model, increasing irradiation changes the equilibrium inclination from about $40^\circ$ to $43^\circ$, reducing its contribution to $R$ by nearly $20\%$ at fixed population fraction. At still higher irradiation the stationary intermediate-$J$ state can lose stability. A stable periodic orbit may then preserve intermediate-$J$ alignment (Section~\ref{sec:oscillatory}). Recurrent returns to thermally wobbling low-$J$ states further reduce alignment.

\section{Rotational disruption: comparing the three regimes}\label{sec:disr}
The relevant quantity for disruption is the \emph{first passage} to $J_{\rm disr}$, not the existence or residence time of a particular attractor. 

\subsection{Disruption histories and orientation-dependent outcomes}
We use the common thermal initial population, infrared damping, and release prescription described in Section~\ref{sec:stochastic}. The illustrative frequency-dependent response is $\delta_m(j)=10/[1+(j/0.2)^2]$. Disruption occurs at the first crossing of $j_{\rm disr}$, whether during the initial evolution or following a release from low $J$.

\begin{figure}[t]
\centering\includegraphics[width=\columnwidth]{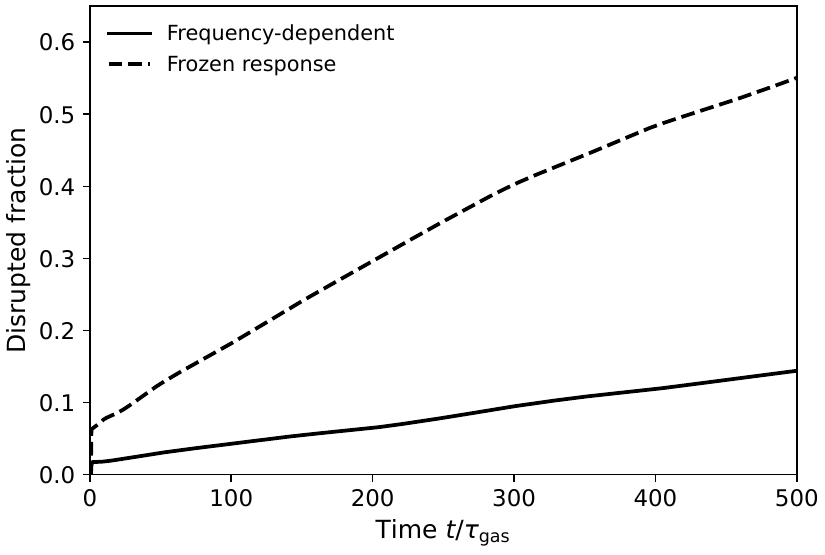}
\caption{Protection against disruption by intermediate-$J$ trapping at $u=1$ and $J_{\rm disr}=0.8J_H$. The initial population, infrared damping and release prescription are those of Section~\ref{sec:stochastic}. Solid and dashed curves show frequency-dependent magnetic relaxation and its frozen control, respectively. Curves are smoothed cumulative first-passage fractions for 1500 histories per model. For the ordinary-RAT control with negligible magnetic relaxation, the adopted illumination geometry has no stable high-$J$ attractor and disruption is negligible.}
\label{fig:ftimemodels}
\end{figure}

The protective effect of intermediate-$J$ trapping is particularly pronounced when the disruption threshold lies well above the intermediate equilibrium but below the nominal RAT spin. Figure~\ref{fig:ftimemodels} illustrates this regime at $u=1$ with $J_{\rm disr}=0.8J_H$. The initial conditions and damping follow Section~\ref{sec:stochastic}. With $J_H=50J_{\rm th}$, this threshold corresponds to $40J_{\rm th}$, illustrating disruption of a weak grain rather than the compact grain used in Figure~\ref{fig:mag}. For a spherical $0.1\,\mu$m grain at $T_g=100$ K and $\rho=3\,\mathrm{g\,cm^{-3}}$, it corresponds to $S_{\max}\sim3\times10^3\,\mathrm{erg\,cm^{-3}}$. The same dimensionless threshold can describe stronger grains at larger nominal RAT spins. After $500\tau_{\rm gas}$, about 14\% of grains disrupt with frequency-dependent magnetic relaxation, compared with 55\% for the frozen response. Intermediate rotation protects grains even when the low-$J$ attractors are long-lived. 

The stationary intermediate equilibrium lies below the adopted disruption threshold, but excursions can still produce disruption. For a periodic attractor, deterministic survival requires the maximum angular momentum along the orbit to remain below the disruption threshold (Section~\ref{sec:oscillatory}). For this geometry ordinary RATs without enhanced magnetic relaxation produce negligible disruption. Magnetic inclusions thus promote disruption relative to ordinary RATs, while the frequency dependence of their response protects grains relative to constant strong magnetic relaxation.

\subsection{Cycling and rare disruption when high-$J$ alignment is unstable}
\label{sec:no-highj-disruption}

When the aligned high-$J$ state is unstable, grains can cycle between low and intermediate rotation. Gas bombardment can occasionally redirect a grain toward disruptive rotation, even though the high-$J$ state cannot retain it. The duration of these cycles depends strongly on the inclusion size.

\subsubsection{Intermediate-$J$ residence and magnetic inclusion size}
\label{sec:cluster-lifetime}

The mean residence time at intermediate rotation depends strongly
on the size of magnetic inclusions. We repeated the two-dimensional
stochastic-escape calculation of Appendix~\ref{app:ir-escape} with
the same RAT geometry and gas-driven fluctuations, varying the
number $N_{\rm cl}$ of iron atoms in each superparamagnetic cluster.
The cluster relaxation time follows the usual thermally activated
response, $\tau_N\propto\exp(N_{\rm cl}T_{\rm act}/T_d)$, where $T_{\rm act}$ is the anisotropy-barrier temperature per magnetic atom and $\tau_N$ is the magnetic-moment reversal time (Appendix~\ref{app:maglaws}).
For this scan we use $T_d=10$ K, the reference low-frequency
magnetic-relaxation strength $\delta_m(0)=10$ at
$N_{\rm cl}=5800$, and $J_{\rm th}/J_H=0.02$.
Table~\ref{tab:cluster-residence} shows the intermediate rotation,
orientation, and mean first-exit time $T_{\rm int}$ from the basin of the attractor, defined in the same way as $T$ in Table~\ref{tab:ir_mfpt}. The diffusion coefficient is the gas-only value ($f_{\rm IR}=0$), allowing the effect of inclusion size to be isolated. The basin boundary is recalculated for each response using the stable manifold of its saddle. The $N_{\rm cl}=5800$ reference model uses $j_c=0.2$ and is identical to the gas-only model in Table~\ref{tab:ir_mfpt}.

\begin{table}[t]
\centering
\caption{Effect of iron-cluster size on the intermediate-$J$ equilibrium and mean first-exit time from its basin, with gas damping alone ($u=1$, $j_{\rm th}=0.02$). Disruption is not imposed in this residence-time calculation.}
\label{tab:cluster-residence}
\setlength{\tabcolsep}{6pt}
\begin{tabular}{rccc}
\hline\hline
$N_{\rm cl}$ & $J_*/J_H$ & $\xi_*$ & $T_{\rm int}/\tau_{\rm damp}$\\
\hline
5000 & 0.77 & $24^\circ$ & $1.7\times10^5$\\
5250 & 0.65 & $31^\circ$ & $3.1\times10^4$\\
5500 & 0.52 & $36^\circ$ & $1.3\times10^3$\\
5750 & 0.42 & $40^\circ$ & $53$\\
5800 & 0.40 & $41^\circ$ & $39$\\
\hline
\end{tabular}
\end{table}

Smaller clusters retain efficient magnetic relaxation at higher rotation frequencies. They can therefore support higher intermediate spins and much longer residence times. The enhancement in Table~\ref{tab:cluster-residence} spans nearly four orders of magnitude. Survival, however, requires the disruption threshold to exceed the angular momentum of the attractor. For the adopted disruption threshold $j_{\rm disr}=0.8$, all the equilibria in the table lie below the threshold. The smallest clusters have the highest intermediate spins and can still be vulnerable to disruptive excursions. The basin residence time is not the survival time.

\subsubsection{Rare excursions to disruptive rotation}

When no stable high-$J$ destination exists, the relevant disruption
time is the accumulated time spent in repeated low- and
intermediate-$J$ episodes before an excursion reaches
$J_{\rm disr}$. If a typical cycle lasts $t_{\rm cyc}$ and a fraction
$p_{\rm disr}$ of cycles reach disruptive rotation, then
\begin{equation}
 t_{\rm disr}\sim\frac{t_{\rm cyc}}{p_{\rm disr}},
 \label{eq:rare-disruption}
\end{equation}
where $t_{\rm cyc}$ includes both low-$J$ waiting and the excursion
to intermediate rotation. A small $p_{\rm disr}$ can make this time much longer than a damping time. The long intermediate-$J$ residence
times in Table~\ref{tab:cluster-residence} can lengthen these cycles
still further. Infrared damping, by reducing stochastic departures,
provides another means of prolonging grain survival.

\subsection{Other cases}
A stable intermediate-$J$ attractor is not a guarantee of mechanical survival. A weak or highly porous aggregate can have a disruption threshold below that of a compact grain, because the centrifugal limit depends on the tensile strength $S_{\max}$ [Equation~(\ref{eq:disr})]. At a rotation frequency $\omega_*$ the stress criterion may be expressed as
\begin{align}
 S_{\max}\lesssim{}& S_*(a,\omega_*)
 =\frac{\rho a^2\omega_*^2}{4}\nonumber\\
 \simeq{}& 7.5\times10^3\,\frac{\mathrm{erg}}{\mathrm{cm}^3}
 \left(\frac{\rho}{3\,\mathrm{g\,cm}^{-3}}\right)
 \left(\frac{a}{0.1\,\mu\mathrm m}\right)^2\nonumber\\
 &\times\left(\frac{\omega_*}{10^7\,\mathrm{s}^{-1}}\right)^2.
 \label{eq:weak-aggregate}
\end{align}
The numerical scaling is illustrative; it does not assign a physical $\omega_*$ to every dimensionless attractor plotted here. An intermediate-$J$ state below the compact-grain disruption threshold can therefore still exceed the threshold of an exceptionally loose aggregate. This provides a natural evolutionary contrast: low-$J$ residence favors the \emph{retention} of fragile aggregates, while inclusion-induced intermediate rotation may remove the loosest members of that population. See \citet{HirashitaYan2009,Ormel2009,HirashitaLi2013} for grain collisions and coagulation.

If ordinary RATs already support stable high-$J$ rotation, magnetic inclusions can accelerate the early spin-up without preventing subsequent high-$J$ capture as their response declines. For instance, for $\alpha=85^\circ$, $\psi=55^\circ$, the mean disruption time among disrupted grains decreases from about $600$ to $350\tau_{\rm damp}$ when frequency-dependent inclusions are added. 

\section{Polarization near illuminating stars}\label{sec:pol}
The loss of magnetically stabilized high-$J$ attractors at high irradiation can reduce polarization toward a star even when grains are not disrupted. For the reference AMO geometry, Equation~(\ref{eq:app_highj_instability}) gives a critical magnetic relaxation parameter $\delta_{m,{\rm crit}}\simeq0.77$. For the reference $0.1\,\mu$m grain, using the analytical relaxation law $\delta_m(j)=10/[1+(j/0.2)^2]$, the aligned high-$J$ state loses stability at $U/U_0\simeq0.69$, corresponding to $J_{\rm RAT}\simeq35J_{\rm th}$. The threshold depends on grain size.

For a star of luminosity $L_\star$, in an optically thin region dominated by direct stellar radiation, the radiation energy density is $L_\star/(4\pi cr^2)$, where $r$ is the distance from the star and $c$ is the speed of light. Thus
\begin{equation}
 \frac{U}{U_0}=\left(\frac{r_0}{r}\right)^2,\qquad
 r_0=\left(\frac{L_\star}{4\pi c U_0 u_{\rm ISRF}}\right)^{1/2},
 \label{eq:pol-distance}
\end{equation}
where $u_{\rm ISRF}$ is the standard interstellar radiation energy density and $U_0$ is defined by the reference RAT spin by the condition $J_H=50J_{\rm th}$. The reference grain loses its high-$J$ attractor inside $r\simeq1.2r_0$. For example, if $L_\star=10^4 L_\odot$ and $u_{\rm ISRF}=8.6\times10^{-13}\,\mathrm{erg\,cm^{-3}}$, then $r_0\simeq3.5\,U_0^{-1/2}$ pc. The value of $U_0$ depends on the gas damping, grain properties, and stellar spectrum.

We calculate the fractions $f_{\rm H}$, $f_{\rm S}$, and $f_{\rm L}$ at high, intermediate, and low $J$ using the model in Appendix~\ref{app:pol-populations}. The fractions satisfy $f_{\rm H}+f_{\rm S}+f_{\rm L}=1$ because grain destruction is excluded from this illustration. Assuming perfect alignment at high $J$ and efficient internal alignment at intermediate $J$, the alignment factor is
\begin{equation}
 R=f_{\rm H}+f_{\rm S}(3\cos^2\xi_*-1)/2,
 \label{eq:pol-population-R}
\end{equation}
where $\xi_*$ is the stationary intermediate-$J$ inclination. This is the stationary-attractor approximation to Equation~(\ref{eq:alignment-regimes}), with $R_{\rm L}=0$ and no explicit periodic population. We use the astrodust cross sections \citep{DraineHensley2021} for oblate grains of axial ratio 1.4 and porosity 0.2, with differential grain number density $dn/da\propto a^{-3.5}$ between 0.005 and $0.35\,\mu$m. The optical model and size distribution are identical for the two magnetic-response prescriptions. The cross-section averaging and population calculation are described in Appendix~\ref{app:pol-populations}.

Figure~\ref{fig:opt} shows the $V$-band extinction polarization and the $850\,\mu$m emission polarization as functions of distance. Approaching the star from $3.2r_0$ to $0.58r_0$ decreases the optical polarization from about 3\% to 0.02\% and the submillimeter polarization from about 6\% to 0.2\% for the frequency-dependent response. The frozen-response model retains much stronger polarization near the star. The decrease results primarily from the loss of high-$J$ alignment, with additional contributions from oblique intermediate-$J$ rotation and the transfer of grains to essentially unaligned low-$J$ states. No change in grain abundance is required.

\begin{figure*}[t]
\centering\includegraphics[width=0.96\textwidth]{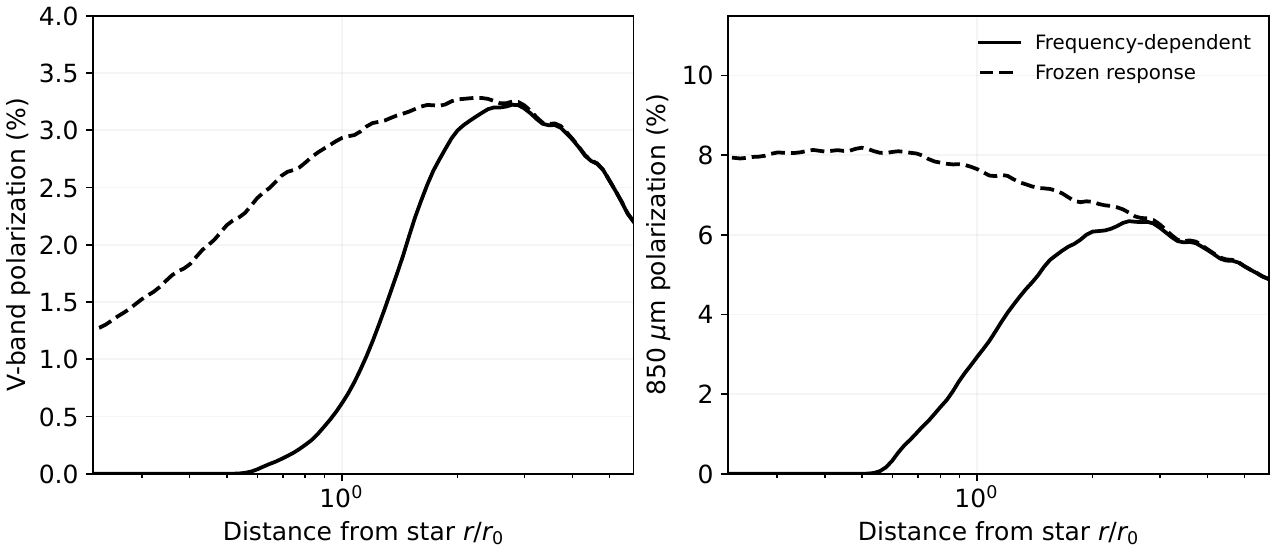}
\caption{Illustrative polarization versus distance from an illuminating star, with $r_0$ defined by Equation~(\ref{eq:pol-distance}). Left: $V$-band extinction polarization at $A_V=1$ mag. Right: optically thin isothermal emission polarization at $850\,\mu$m. Solid and dashed curves show frequency-dependent and frozen magnetic relaxation, respectively. At each distance the local population evolves for $1000\tau_{\rm ref}$, where $\tau_{\rm ref}=\tau_{\rm gas}(0.1\,\mu\mathrm{m})$. The magnetic field is in the plane of the sky and the grain-size distribution is fixed. Curves are lightly smoothed.}\label{fig:opt}
\end{figure*}

The different radial trends of the frozen-response curves arise from grain-size-dependent alignment and optical weighting. At stronger irradiation, smaller grains attain suprathermal rotation, whereas larger grains are captured by high-$J$ attractors less efficiently within the adopted exposure time. This reduces the $V$-band polarization, which is particularly sensitive to grains of $0.1$--$0.2\,\mu$m. At $850\,\mu$m, the increased alignment of smaller grains largely compensates for the reduced contribution of larger grains.

Spatially resolved measurements of polarization and extinction around stars, such as HD~97300 or $\gamma$~Cas, can test this prediction \citep{Andersson2011}. For material lying approximately in the plane of the sky, projected stellar separation approximates $r$. More generally, observations integrate through material at different distances. A decrease of polarization toward the star without a corresponding change in the extinction curve would favor a change in alignment over grain destruction, provided variations in magnetic-field geometry are accounted for. 

\section{Discussion}\label{sec:disc}

Our present work is related to both radiative and mechanical torque effects. 
Frequency-dependent dissipation may similarly alter MET alignment and disruption of drifting grains \citep{LH2007met,HoangChoLazarian2018,LH2021}. As RATs field is more developed, we focus on RATs, while a detailed discussion of METs is given in a companion paper (Lazarian 2026).

\subsection{Big picture: alignment, grain growth, and disruption}
\label{sec:big_picture}

The existing RAT theory predicts that for ordinary paramagnetic grains the efficient RAT alignment at high-$J$ takes place for a range of angle between the radiation and magnetic field. This range depends on the grain shape. Grains in the vicinity of high-$J$ attractors can rotate fast enough to induce grain disruption. On the contrary, the alignment at low-$J$ attractors is very low. 

If grains have both low- and high-$J$ attractors, initially most of the grains from an ensemble with uniformly distributed angular momenta directions move to the low-$J$ attractors. Then, slowly, over many gas damping times, grains diffuse from low-$J$ to a more stable high-$J$ attractor. This time-scale is extended for high radiative flux.

In the absence of high-$J$ attractor, 
gas collisions can occasionally launch a trajectory close enough to a high-$J$ repeller to cross $J_{\rm disr}$, but stronger radiative forcing can decrease the frequency of these exceptional routes. During extended intervals near low $J$, grains avoid centrifugal breakup and may coagulate into loose aggregates, provided grain--grain impacts are sufficiently gentle and adhesive. 

Adding efficient magnetic inclusions extends the range of angles and grain shapes for which high-$J$ attractors exist.
This should be separated from an incorrect understanding of magnetic RAT (MRAT) as a process in which the alignment is due to magnetic dissipation with RATs providing a steady isotropic spin-up of grains. RATs are highly anisotropic and impose their own alignment. Torques arising from magnetic dissipation tip the balance, and modify the RAT alignment. This way, magnetic dissipation enhances alignment and provides condition for RAT disruption (RATD) process. 

For sufficiently strong magnetic dissipation the parameter space over which the MRAT acts is significant (see Figure~\ref{fig:amo-regimes}) which makes possible essentially all grains with magnetic inclusions to be perfectly aligned fast and also fast disrupted if the radiation is strong enough. 

The present paper challenges this simplistic picture. Figure \ref{fig:mag} shows that, for sufficiently fast rotation, magnetic inclusions stop inducing additional dissipation and the dissipative response $K(\omega)$ approaches the ordinary paramagnetic value. It is important that the corresponding rotational rates are lower than the disruption rate of solid grains. As a result, for a significant parameter space, grains reach intermediate-$J$ attractors at which they are partially aligned and protected against disruption when their rotation remains below the disruption threshold. This is a significant new element of grain dynamics that was missed by the earlier studies. 

Grains in the intermediate-$J$ state are subject to diffusion induced by gaseous bombardment. This diffusion can move some grains to low-$J$ attractors, from which they can be released and recaptured at intermediate $J$. Occasionally, grains can undergo trips in the vicinity of the high-$J$ repeller and be disrupted if the radiation is strong enough. Table~\ref{tab:environment-survival} illustrates that intermediate-$J$ residence alone can exceed the available exposure time in some astrophysical environments. 

In general, our present study extends the picture of anisotropic alignment and disruption suggested in \cite{LH2021} for paramagnetic grains to grains with magnetic inclusions. It also demonstrates that the decrease of polarization with the intensity of radiation does not necessarily mean grain disruption, but can arise from grain alignment in intermediate-$J$ attractors. 

The existence of intermediate-$J$ attractors must be accounted for the processes of disruption, coagulation of grains and the evolution of grain mantles. This calls for detailed study of the new phenomena with the focus on the transition from stationary intermediate-$J$ state to the oscillatory one. The latter can have lower residence time, as our preliminary study shows. 

\subsection{Survival and alignment in typical environments}\label{sec:finite-time}

In finite-lived astrophysical environments, grain survival depends on the time $t_{\rm disr}$ to reach $J_{\rm disr}$ relative to the grain radiation exposure time $t_{\rm exp}$ within the environment. Long intermediate-$J$ residence can protect grains against rotational disruption if the intermediate rotation rate remains below the disruption threshold.

For a grain of radius $a$, the gas damping time is approximately \citep{Andersson2015, Hoang2020Review}
\begin{equation}
\begin{split}
\tau_{\rm gas}\simeq{}&8.7\times10^4\,{\rm yr}\,
\left(\frac{a}{0.1\,\mu{\rm m}}\right)
\left(\frac{\rho}{3\,{\rm g\,cm^{-3}}}\right)\\
&\times\left(\frac{30\,{\rm cm^{-3}}}{n_{\rm H}}\right)
\left(\frac{100\,{\rm K}}{T_g}\right)^{1/2},
\end{split}
\label{eq:environment-gas-damp}
\end{equation}
Infrared emission can shorten this time to $\tau_{\rm damp}=\tau_{\rm gas}/(1+F_{\rm IR})$, with $F_{\rm IR}\simeq0.4U^{2/3}(0.1\,\mu{\rm m}/a)(30\,\mathrm{cm}^{-3}/n_{\rm H})(100\,\mathrm K/T_g)^{1/2}$ \citep{Hoang2020Review}.
The ratio $F_{\rm IR}$ measures damping, while stochastic recoil is set by the angular momentum carried by emitted photons. These distinct contributions enter the residence-time calculation in Appendix~\ref{app:ir-escape}.

The infrared contribution becomes comparable to gas damping when
\begin{equation}
n_{\rm H}\lesssim n_{\rm IR}\simeq 12\,U^{2/3}
\left(\frac{0.1\,\mu{\rm m}}{a}\right)
\left(\frac{100\,{\rm K}}{T_g}\right)^{1/2}
{\rm cm^{-3}}.
\label{eq:ir-density-threshold}
\end{equation}
The infrared fraction is $f_{\rm IR}=F_{\rm IR}/(1+F_{\rm IR})$. Table~\ref{tab:environment-survival} compares the mean intermediate-$J$ residence time $t_{\rm res}$, estimated by logarithmic interpolation of the dimensionless mean first-exit times in Table~\ref{tab:ir_mfpt} and multiplication by the local $\tau_{\rm damp}$, with the available exposure time. 

\begin{table*}[t]
\centering
\caption{Environmental conditions and intermediate-$J$ residence times relevant to grain survival. The estimates assume the existence of the reference intermediate-$J$ attractor and adopt $a=0.1\,\mu$m and $\rho=3\,\mathrm{g\,cm^{-3}}$ and the reference torque geometry of Table~\ref{tab:ir_mfpt}. The residence time is not the disruption time. Exposure times represent the planetary-nebula lifetime or a local outflow crossing time.}
\label{tab:environment-survival}
\setlength{\tabcolsep}{5pt}
\begin{tabular}{lrrrrrrr}
\toprule
Environment & $n_{\rm H}$ & $T_g$ & $U$ & $f_{\rm IR}$ & $\tau_{\rm damp}$ & $t_{\rm res}$ & $t_{\rm exp}$ \\
 & (cm$^{-3}$) & (K) & & & (yr) & (yr) & (yr) \\
\midrule
Planetary nebula, dense & $10^4$ & $10^4$ & $10^4$ & 0.05 & 25 & $\sim10^3$ & $2\times10^4$ \\
Planetary nebula, diffuse & $10^3$ & $10^4$ & $10^4$ & 0.36 & 170 & $2.3\times10^4$ & $2\times10^4$ \\
Circumstellar outflow, dense & $10^7$ & $10^3$ & $10^6$ & 0.004 & 0.08 & $\sim3$ & 47 \\
Circumstellar outflow, diffuse & $10^5$ & $10^3$ & $10^6$ & 0.28 & 6 & 550 & 47 \\
\bottomrule
\end{tabular}
\end{table*}

In planetary nebulae, the intermediate-$J$ residence time $t_{\rm res}$ can approach or exceed the observable nebular lifetime of $t_{\rm exp}\sim2\times10^4$ yr \citep{Jacob2013PN}. This occurs in the less dense illuminated regions where infrared damping reduces stochastic escape. The resulting suppression of high-$J$ alignment and disruption is relevant to the survival of micron-size silicates observed in NGC~6302 \citep{Matsuura2025NGC6302}.

In circumstellar outflows, a grain moving at $10\,\mathrm{km\,s^{-1}}$ crosses 100 AU in 47 yr. In the lower-density example, the intermediate-$J$ residence time exceeds this exposure time by about an order of magnitude. For a freely expanding wind with $n_{\rm H}\propto r^{-2}$ and $U\propto r^{-2}$, $F_{\rm IR}\propto r^{2/3}$ at fixed gas temperature, allowing a radial transition in alignment and disruption. Infrared damping is less important for the larger $0.3$--$0.5\,\mu$m grains inferred in some stellar outflows \citep{Aronson2017AGB,Scicluna2015VY}.

Interstellar grains entering the heliosphere encounter solar-wind electromagnetic forces during a transit lasting decades \citep{NgGuHoang2025Helio}. Intermediate rotation may contribute to their survival. Stratospheric dust associated with interstellar silicates contains iron--nickel metal and iron-rich sulfide inclusions \citep{Bradley1994GEMS}. This association is supported by the isotopic evidence \citep{Messenger2000MolecularCloud}. 

Variable illumination from novae or supernovae offers a complementary time-domain test \citep{LH2019,LH2021,Hoang2019}. An early change in $p(\lambda)$ or $\lambda_{\max}$ without corresponding extinction evolution would favor a change in alignment over disruption.

\subsection{Grain size and the frequency of magnetic relaxation}\label{sec:sizedep}
The response of an inclusion must be evaluated at the grain's actual rotation frequency. For a spherical grain of density $\rho=3\,{\rm g\,cm^{-3}}$, the thermal angular velocity is
\begin{equation}
\begin{aligned}
\omega_{\rm th}&=\left(\frac{k_{\rm B}T_g}{I}\right)^{1/2}\\
&\simeq1.7\times10^5\,{\rm s^{-1}}
\left(\frac{T_g}{100\,{\rm K}}\right)^{1/2}\\
&\quad\times\left(\frac{\rho}{3\,{\rm g\,cm^{-3}}}\right)^{-1/2}
\left(\frac{a}{0.1\,\mu{\rm m}}\right)^{-5/2}.
\end{aligned}
\label{eq:omega-size}
\end{equation}
At $J/J_{\rm th}=100$ and $T_g=100$ K, grains of $0.3$ and $10\,\mu$m rotate at approximately $10^6$ and $170\,\mathrm{s^{-1}}$, respectively. Thus they probe very different frequencies of the same inclusion response. When $\omega\tau_{\rm resp}\gg1$, a slowly fluctuating cluster contributes little dissipation \citep{LH2019}. Circumstellar grains around $0.3$--$1\,\mu$m can sample either response regime depending on the inclusion relaxation time $\tau_{\rm resp}$.  

Inclusion size must be distinguished from grain radius. For a power-law distribution of superparamagnetic clusters, \citet{Hoang2026SPI} found that the clusters contributing most strongly at rotation frequency $\omega$ satisfy approximately
\begin{equation}
N_{\rm res}(\omega,T_d)=\frac{T_d}{T_{\rm act}}\ln\!\left(\frac{\nu_0}{\omega}\right),
\label{eq:nres}
\end{equation}
where $T_{\rm act}$ denotes the anisotropy barrier per magnetic unit and $\nu_0$ the attempt frequency. With the iron parameters of \citet{Hoang2026SPI}, $T_d=10$ K gives $N_{\rm res}\simeq8\times10^3$ (cluster radius $\sim2.9$ nm) at $\omega=10^5\,\mathrm{s^{-1}}$ and $N_{\rm res}\simeq1.5\times10^4$ ($\sim3.5$ nm) at $10^2\,\mathrm{s^{-1}}$. 

For a single inclusion class, magnetic relaxation can weaken between low and high spin when its characteristic time lies within
\begin{equation}
\omega_{\rm high}^{-1}\ \lesssim\ \tau_N(T_d)\ \lesssim\ \omega_{\rm low}^{-1},
\label{eq:selection-window}
\end{equation}
where $\omega_{\rm low}$ and $\omega_{\rm high}$ bracket the relevant spin states. The resulting attractors depend on the torque geometry, inclusion abundance, precession, and stochastic fluctuations.

If we assume a broad distribution of inclusion sizes as in \citep{Hoang2026SPI}, the processes of disruption will favor to survival of grains with more stable intermediate-$J$ attractors. This is an interesting possibility to be explored elsewhere.

Our phase portraits assume rapid internal alignment of the maximum-inertia axis with $\boldsymbol J$. This differs from external magnetic relaxation of $\boldsymbol J$ toward $\boldsymbol B$. Barnett and nuclear-spin dissipation depend on body-frame wobble frequencies, which need not equal $J/I$ \citep{LazarianDraine1999,LH2019}.
Inelastic dissipation provides another channel: wobble stresses at twice the precession frequency can dominate Barnett damping in oblate or aggregated grains \citep{LazarianEfroimsky1999}. Higher harmonics, triaxial motion, and slow relaxation near separatrices were explored by \citet{EfroimskyLazarian2000,Efroimsky2000,Efroimsky2001,EfroimskyLazarianSidorenko2002}. Therefore loss of magnetic internal relaxation at rapid spin need not imply inefficient internal alignment.

\section{Summary}\label{sec:summary}
\begin{enumerate}
\item For RATs at sufficiently strong illumination, frequency-dependent magnetic dissipation can induce suprathermal intermediate-$J$ attractors. Stronger illumination can therefore decrease grain alignment and rotational disruption. Similar effects are expected for METs at sufficiently large grain--gas drift velocities.

\item Intermediate-$J$ attractors can correspond to stationary or oscillatory grain rotation. The latter arises from the feedback between radiative spin-up, grain inclination, and frequency-dependent magnetic relaxation. Grains undergoing oscillatory rotation can remain appreciably aligned while avoiding disruption if the maximum angular momentum along the cycle remains below the disruption threshold. Systematic body-fixed torques, including Purcell and infrared emission torques, can also induce transitions to oscillatory rotation.

\item Infrared emission damps rotational departures that can carry grains away from intermediate-$J$ attractors, while its photon recoil produces weaker stochastic angular-momentum kicks per unit damping than gas collisions. Both effects favor residence near low and intermediate-J attractors, contributing to longer waiting time for achieving grain disruption. 
\item Actual disruption time must be compared with the exposure duration. In planetary nebulae, circumstellar winds and other environments, the trapping induced by stationary intermediate-$J$ attractors can be longer than the typical residence time of dust in these environments. 
\item Broad distributions of inclusion sizes smooth the magnetic response, while differences among grains can favor the survival of particular grain and inclusion sizes. 
\item When high-$J$ alignment becomes unstable at rapid rotation,
grains can survive through long intervals at low and intermediate
$J$, with only rare excursions reaching disruptive rotation.
The inclusion size strongly controls the duration: in our reference
calculation, reducing the iron-cluster size from 5800 to 5000 atoms
increases intermediate-$J$ residence by nearly four orders of
magnitude.
\item Grains can
lose their MRAT alignment advantage near strong radiation
sources, decreasing the polarization with increasing radiation intensity. The duration of the intermediate-$J$ state can exceed the dust residence time in many environments.
\end{enumerate}

\appendix
\section{Dynamical equations and stability of the intermediate-$J$ state}\label{app:dynamics}
For rapid magnetic-field-dominated precession, the secular evolution of the angle $\xi$ between $\boldsymbol J$ and $\boldsymbol B$ and of the angular momentum $J$ is
\begin{align}
\frac{d\xi}{dt}&=\frac{M\langle F(\xi,\psi)\rangle}{J}-\frac{\sin\xi\cos\xi}{\tau_{\rm mag}(J)}, \label{eq:app_phase_xi}\\
\frac{dJ}{dt}&=M\langle H(\xi,\psi)\rangle-\frac{J}{\tau_{\rm damp}}-\frac{J\sin^2\xi}{\tau_{\rm mag}(J)}. \label{eq:app_phase_J}
\end{align}
Here $M=uM_0$, and $\tau_{\rm mag}^{-1}(J)\propto K(J/I)B^2$. Near thermal rotation, internal fluctuations and gas impacts supplement these deterministic equations.

At $\xi=0$, magnetic braking vanishes and $J_{\rm RAT}=M\langle H(0,\psi)\rangle\tau_{\rm damp}$. For $\langle F(0)\rangle=0$, this equilibrium is angularly unstable when
\begin{equation}
\frac{M}{J_{\rm RAT}}\left.\frac{d\langle F\rangle}{d\xi}\right|_{0}>\frac{1}{\tau_{\rm mag}(J_{\rm RAT})}.
\label{eq:app_highj_instability}
\end{equation}
A stationary intermediate-$J$ attractor requires a stable off-axis equilibrium. With $j=J/J_H$ and $u=U/U_0$, its fixed-point conditions are
\begin{equation}
\frac{u f(\xi)}{j}=\delta_m(j)\sin\xi\cos\xi,\qquad
u h(\xi)=j\left[1+\delta_m(j)\sin^2\xi\right],
\label{eq:app_balances}
\end{equation}
Here $\delta_m(j)=\tau_{\rm damp}/\tau_{\rm mag}(j)$, and $f,h$ are the normalized RAT components. Primes denote derivatives with respect to $\xi$ for $f,h$ and $j$ for $\delta_m$. Along an off-axis branch, angular balance gives
\begin{equation}
 d\ln u=\left(1+\frac{d\ln \delta_m}{d\ln j}\right)d\ln j+
 \left(2\cot 2\xi-\frac{f'(\xi)}{f(\xi)}\right)d\xi.
\label{eq:app_branch_sensitivity}
\end{equation}
The inclination response to irradiation thus depends on the torque slopes and magnetic-response law. At $(j_*,\xi_*)$, with $s=\sin\xi_*$ and $c=\cos\xi_*$, the Jacobian is
\begin{equation}
{\cal A}=\begin{pmatrix}
 -1-\delta_m s^2-j\delta_m's^2 & u h'-2j\delta_m sc\\
 -u f/j^2-\delta_m'sc & u f'/j-\delta_m(c^2-s^2)
\end{pmatrix}_{(j_*,\xi_*)}.
\label{eq:app_jacobian}
\end{equation}
Stability requires ${\rm tr}\,{\cal A}<0$ and $\det {\cal A}>0$. Complex eigenvalues with negative real parts give an inward spiral. The stationary attractor ceases to be stable when the fixed point disappears or loses stability. A crossing of ${\rm tr}\,{\cal A}=0$ with $\det {\cal A}>0$ and nonzero crossing slope indicates a possible Hopf bifurcation. The periodic attractor in Section~\ref{sec:oscillatory} is identified by trajectory integration.

\section{Magnetic-response laws used in calculations}\label{app:maglaws}
The frequency-dependent magnetic-relaxation parameter entering the dynamical equations can be written as
\begin{equation}
\delta_m(j)=\frac{5\tau_{\rm damp}B^2}{2\rho a^2}\,K(j\omega_H,T_d),
\label{eq:app_delta_m}
\end{equation}
where the factor $5/(2\rho a^2)$ assumes the inertia-to-volume ratio of a homogeneous sphere, $\omega_H=J_H/I$, and $K(\omega)=\chi_2(\omega)/\omega$ is the dissipative magnetic coefficient. For ordinary paramagnetic material we used
\begin{equation}
K_{\rm SS}(\omega)=\frac{K_{\rm SS}(0)}{1+\omega^2\tau_{\rm SS}^2},
\label{eq:app_kss}
\end{equation}
where $K_{\rm SS}(0)$ is the zero-frequency spin--spin coefficient and $\tau_{\rm SS}$ its relaxation time. For the independent-cluster superparamagnetic prescription we used
\begin{equation}
K_{\rm SP}(\omega)=K_{\rm SS}(\omega)+\frac{\chi_{\rm SP}(0)\tau_{\rm fluct}}{1+\omega^2\tau_{\rm fluct}^2},
\label{eq:app_ksp}
\end{equation}

Here $\chi_{\rm SP}(0)$ is the static cluster susceptibility and $\tau_{\rm fluct}$ its magnetic-moment fluctuation time, identified with $\tau_N$ below.

\paragraph{Superparamagnetic cluster size and the cutoff frequency.}
At fixed magnetic-atom abundance, the cluster number density scales as $N_{\rm cl}^{-1}$ and each moment as $N_{\rm cl}$. Thus $\chi_{\rm SP}(0)\propto N_{\rm cl}/T_d$. N\'eel reversal gives
\begin{equation}
 \tau_N=\tau_0\exp\!\left(\frac{N_{\rm cl}T_{\rm act}}{T_d}\right),
 \qquad \omega_c\simeq\tau_N^{-1},
 \label{eq:app_neel_size}
\end{equation}
where $T_{\rm act}$ is an anisotropy-barrier temperature per magnetic
atom and $\tau_0$ is the microscopic attempt time. The cluster
contribution to the dissipative coefficient is
\begin{equation}
 K_{\rm SP,cl}(\omega,N_{\rm cl})=
 \frac{\chi_{\rm SP}(0,N_{\rm cl})\tau_N}
 {1+\omega^2\tau_N^2}.
 \label{eq:app_cluster_response}
\end{equation}
Consequently,
\begin{equation}
 K_{\rm SP,cl}(0)\propto N_{\rm cl}\tau_N,
 \qquad
 K_{\rm SP,cl}(\omega\tau_N\gg1)\propto
 \frac{N_{\rm cl}}{\omega^2\tau_N}.
 \label{eq:app_cluster_limits}
\end{equation}
At fixed $T_d$, larger clusters enhance low-frequency dissipation but lower the turnover frequency and can dissipate less at high rotation rates. In the $N_{\rm cl}=5000$--5800 scan, $\delta_m(0)$ increases from 3.57 to 10 while $j_c=\omega_c/\omega_H$ (the normalized turnover angular momentum) decreases from 0.481 to 0.2. A distribution of cluster sizes broadens the turnover \citep{Hoang2026SPI}.

\paragraph{Spin--lattice response enabled by magnetic inclusions.}
Local magnetic fields from ferromagnetic and ferrimagnetic inclusions enable enhanced spin--lattice relaxation in the paramagnetic matrix \citep{LH2019}. Retaining the ordinary spin--spin contribution, we adopt
\begin{equation}
K_{\rm SL}(\omega)=K_{\rm SS}(\omega)+
\frac{\Delta K_{\rm SL}(0)}{1+\omega^2\tau_{\rm SL}^2},
\label{eq:app_ksl}
\end{equation}
Here $\Delta K_{\rm SL}(0)>0$ is the additional low-frequency dissipation and $\tau_{\rm SL}$ its relaxation time. At high frequency the enhancement falls as $\omega^{-2}$, leaving the paramagnetic baseline. More generally, independent dissipative channels give
\begin{equation}
 K(\omega)=K_{\rm SS}(\omega)+
 \sum_q\frac{A_q\tau_q}{1+\omega^2\tau_q^2},
 \label{eq:app_magnetic_channels}
\end{equation}
Here $A_q$ and $\tau_q$ are the amplitude and relaxation time of channel $q$. Coupled relaxation of the same magnetization instead requires a coupled susceptibility.

For Figure~\ref{fig:mag}, we use $K_{\rm SS}(0)=1.2\times10^{-13}$~s,
$\tau_{\rm SS}=10^{-12}$~s, and
$\Delta K_{\rm SL}(0)=3.6\times10^{-9}$~s with
$\tau_{\rm SL}=10^{-6}$~s. The illustrative cluster contributions
for $N_{\rm cl}=300$, 1000, and 3000 have zero-frequency amplitudes
of $7.5\times10^{-10}$, $2.8\times10^{-7}$, and $0.5$~s and
turnover frequencies of $1.7\times10^8$, $1.2\times10^6$, and
$1.5$~s$^{-1}$, respectively. These parameters illustrate the
response shapes rather than a common calibrated grain composition.
The numerical benchmarks for superparamagnetic grains use the
separate prescriptions specified in the corresponding sections.

Gas bombardment is included through angular-momentum diffusion:
\begin{equation}
 d\boldsymbol j=\left[\dot j\,\hat{\boldsymbol j}+j\dot\xi\,\hat{\boldsymbol e}_{\xi}\right]dt'
 +\sqrt{2}\,j_{\rm th}\,d\boldsymbol W_{t'},
\label{eq:app_stochastic}
\end{equation}
For the gas-only stochastic benchmarks, $\tau_{\rm damp}=\tau_{\rm gas}$. Here $\boldsymbol j=\boldsymbol J/J_H$, $\hat{\boldsymbol j}=\hat{\boldsymbol J}$, $\boldsymbol W_{t'}$ is a three-dimensional standard Wiener process, and $j_{\rm th}=\sqrt{k_{\rm B}T_g/I}/\omega_H$ sets the strength of
the random kicks and $t'=t/\tau_{\rm damp}$.

\section{Gas and infrared damping versus stochastic escape}
\label{app:ir-escape}
We use the total rotational damping rate $\gamma=\tau_{\rm damp}^{-1}$ and its infrared fraction $f_{\rm IR}=F_{\rm IR}/(1+F_{\rm IR})$, defined in Section~\ref{sec:physics} [Equation~(\ref{eq:damping-definition})].
The diffusion tensors add independently,
$\boldsymbol{\mathsf D}=\boldsymbol{\mathsf D}_{\rm g}
+\boldsymbol{\mathsf D}_{\rm IR}$, with the convention
$\langle dJ_i\,dJ_j\rangle=2\mathsf D_{ij}\,dt$.
Infrared photons and colliding gas particles have different
angular-momentum increments and statistics.

We use the two-dimensional deterministic drift $(a_j,a_\xi)=(\dot j,\dot\xi)$ of Equations~(\ref{eq:dj}) and (\ref{eq:dxi}), with the tabulated AMO torque functions and the reference magnetic response $\delta_m(j)=10/[1+(j/0.2)^2]$.
At $u=1$, these equations have a stationary intermediate-$J$ attractor at
$(j_*,\xi_*)=(0.40,41^\circ)$ and a saddle at
$(j_s,\xi_s)=(0.11,50^\circ)$. Small displacements from the attractor decay while the grain
spirals back toward it. The calculated decay time is about
$10.2\tau_{\rm damp}$ (with an oscillation frequency
of about $1.4/\tau_{\rm damp}$).

The escape boundary is traced numerically along the stable manifold of the saddle point.
It separates grains returning to the intermediate-$J$ attractor
from grains evolving away from it. Its closest
approach to the attractor corresponds to an angular-momentum
separation of approximately $0.10J_H$. Unlike an assumed interval
$|J-J_*|<b$, this boundary allows a grain to leave the attractor
through coupled changes of $J$ and $\xi$.

The resulting escape boundary is shown in Figure~\ref{fig:ir_basin}. The residence times calculated here apply to the stationary attractor, not the periodic attractor in Section~\ref{sec:oscillatory}.

We now change the infrared fraction $f_{\rm IR}$ while keeping the
same deterministic torques and total damping rate. The relative
angular-momentum diffusion coefficient is
\begin{equation}
 D=j_{\rm th}^{\,2}\,r,\qquad
 r=1-(1-\epsilon)f_{\rm IR},\qquad \epsilon=0.01,
 \label{eq:ir_diffusion_ratio}
\end{equation}
where $j_{\rm th}=0.02$ is the noise amplitude in the original
stochastic calculations. The factor $\epsilon$ accounts for the
smaller random angular-momentum increments per unit damping from
infrared emission than from colliding gas atoms. The random torque
acts on all three components of the angular momentum, as in
Equation~(\ref{eq:app_stochastic}). We include gas excitation and infrared recoil but neglect fluctuations associated with magnetic dissipation. The residence times therefore describe this adopted diffusion model.

The mean residence time $T(j,\xi)$ until the grain first leaves the
region surrounding the intermediate-$J$ attractor can be found directly from
\begin{equation}
\begin{split}
-1={}&a_j\,\frac{\partial T}{\partial j}
       +a_\xi\,\frac{\partial T}{\partial\xi}\\
 &+D\left[\frac{\partial^2T}{\partial j^2}
       +\frac{2}{j}\frac{\partial T}{\partial j}
       +\frac{1}{j^2}\left(
          \frac{\partial^2T}{\partial\xi^2}
          +\cot\xi\frac{\partial T}{\partial\xi}\right)\right],
\end{split}
\label{eq:ir_backward}
\end{equation}
with $T=0$ at the escape boundary shown in
Figure~\ref{fig:ir_basin}. We also stop trajectories
that reach the low-$J$ cutoff $j=0.025$. This equation retains
both angular and radial diffusion and uses the actual restoring
motion around the attractor. Table~\ref{tab:ir_mfpt} gives the mean time evaluated at the intermediate attractor in units of $\tau_{\rm damp}$; $T_0$ denotes the value for gas damping alone.

\paragraph{Stochastic benchmarks for Section~\ref{sec:AMO}.}
For the stochastic histories in Figure~\ref{fig:ftimemodels}, we integrate the Cartesian angular momentum with Gaussian increments of component variance $2j_{\rm th}^2(1-0.99f_{\rm IR})\Delta t'$. The time step is $\Delta t'=0.04\min(1,j/u)$, and the low-$J$ core is treated by the release prescription in Section~\ref{sec:stochastic}. We use 1500 histories per curve and check the results with 600 additional histories at half the time step. The endpoint fractions agree within four percentage points. For display we interpolate cumulative fractions monotonically through a coarse time grid, retaining the rapid initial evolution and the final fractions. The other numerical benchmarks below use their separately specified parameters.

For the same reference magnetic response, we separately measured the first passage to $j<0.025$ from the intermediate attractor using 350 histories per noise amplitude, followed for $1000\tau_{\rm damp}$. The median return times for $J_{\rm th}/J_H=0.04,0.03,0.02$ are $8.2$, $15.4$, and $49.7\tau_{\rm damp}$. At $0.015$, the median among returning trajectories is $193\tau_{\rm damp}$ and $4.3\%$ have not returned by the end of the integration. At $0.01$, $96.6\%$ have not returned. The perturbation decay time rises from about $10\tau_{\rm damp}$ at $u=1$ to $56\tau_{\rm damp}$ near $u=1.04$.

\paragraph{Cluster-size residence calculation.}
The smaller-cluster models use the activated-response parameters $(j_c,\delta_m(0))=(0.481,3.56656)$, $(0.365062,4.93421)$, $(0.277069,6.81083)$, and $(0.210285,9.38176)$ for $N_{\rm cl}=5000$, 5250, 5500, and 5750, respectively. For the 5800-atom reference model we adopt $(j_c,\delta_m(0))=(0.2,10)$, exactly as in Table~\ref{tab:ir_mfpt}. These discrete response models follow the size dependence of the activated response in Appendix~\ref{app:maglaws}, with the reference endpoint set to the common benchmark. For each model the intermediate equilibrium and saddle determine the basin boundary, and Equation~(\ref{eq:ir_backward}) gives the mean first-exit time with gas-only diffusion.

\section{Illustrative polarization model}\label{app:pol-populations}
We hold the gas conditions, grain temperature, magnetic field, radiation spectrum, anisotropy, and AMO geometry fixed. For $a_0=0.1\,\mu$m, we adopt $J_H=50J_{\rm th}(a_0)$ at $U_0$. With gas damping alone, $I\propto a^5$, $\tau_{\rm gas}\propto a$, and the approximate RAT efficiency $Q(a)=\min[(a/0.25\,\mu{\rm m})^3,1]$ \citep{LH07}, the characteristic spin frequency and magnetic relaxation parameter are
\begin{align}
 \omega_{\rm RAT}/\omega_H&=u[Q(a)/Q(a_0)](a/a_0)^{-2},\\
 \delta_m(a,\omega)&=\frac{10(a_0/a)}{1+(\omega/0.2\omega_H)^2},
\end{align}
where $\omega_{\rm RAT}=J_{\rm RAT}/I(a)$, $\omega_H=J_H/I(a_0)$, and $u=U/U_0$. The frozen control omits the frequency-dependent denominator. We neglect the ordinary paramagnetic contribution in this strongly enhanced reference model.

The population fractions in Equation~(\ref{eq:pol-population-R}) are estimated from deterministic trajectories released at $3J_{\rm th}$ with isotropic directions. We classify capture by stable suprathermal equilibria ($J\geq5J_{\rm th}$) and return to low $J$. Starting with all grains at low $J$, we use the capture fractions per release as branching probabilities, with mean release interval $t_{\rm L}=100\tau_{\rm gas}/\min[5,\max(1,J_{\rm RAT}/50J_{\rm th})]$ and an illustrative mean intermediate-$J$ episode duration of $72\tau_{\rm gas}$. This prescribed duration in the population model is distinct from the mean first exit through the stationary basin boundary in Table~\ref{tab:ir_mfpt}. Stable high-$J$ capture is treated as absorbing. The local populations evolve for the exposure time in Figure~\ref{fig:opt}.

We integrate the tabulated astrodust extinction and absorption cross sections (zero metallic-iron fraction) over the adopted size distribution. For cross sections $C_1$ (propagation along the symmetry axis), $C_2$ and $C_3$ (perpendicular propagation, electric vector parallel and perpendicular to the axis), we approximate the polarized and mean cross sections by $C_{\rm pol}=(C_3-C_2)/2$ and $C_{\rm mean}=(1-R)(C_1+C_2+C_3)/3+R(C_2+C_3)/2$. The extinction polarization is $\tanh(\int nR C_{\rm pol,ext}\,da\,ds)$, normalized to $A_V=1$ mag. The optically thin isothermal emission polarization is the ratio of the size-integrated $nR C_{\rm pol,abs}$ to $nC_{\rm mean,abs}$ in magnitude. Here $n=dn/da$ is the differential grain number density and $s$ is the line-of-sight coordinate. The fixed grain abundance isolates the effect of alignment from disruption.

\bibliographystyle{aasjournal}
\bibliography{references}

\end{document}